\documentclass[useAMS,usenatbib,onecolumn]{mnras}
\usepackage{graphicx,rotating}
\usepackage{longtable}

 \title[A reference sample of face-on bulgeless galaxies]{A reference sample of face-on bulgeless galaxies} 

 \author[I. D. Karachentsev,   V. E. Karachentseva]{I. D. Karachentsev$^1$\thanks{E-mail: ikar@sao.ru}, V. E. Karachentseva$^2$ \\
$^1$Special Astrophysical Observatory, Russian Academy of Sciences, N.Arkhyz, 369167 Russia\\ 
$^2$Main Astronomical Observatory of National Academy of Sciences of Ukraine, Kiev 03143 Ukraine} 

\begin{document}
\date{Accepted XXX. Received XXX; in original form XXX}

\pagerange{\pageref{firstpage}--\pageref{lastpage}} \pubyear{XXX}

\maketitle

\label{firstpage}

\begin{abstract}
 We present a list of 220 face-on, almost bulgeless galaxies assumed to be counterparts to the objects from the Reference Flat Galaxy Catalog (RFGC). We selected the Sc, Scd and Sd-type galaxies according to their apparent axial ratio $log(r_{25}) < 0.05$ and major standard angular diameter $log(d_{25}) > 0.90$ as defined in HyperLEDA. The sample objects are restricted by the radial velocity $V_{LG} < 10000$ km/s and a declination of above --30 deg. The morphological composition of our sample is quite similar to that of RFGC. We notice the following common properties of face-on bulgeless galaxies. About half of them have bar-like structures occurring in the whole range of the absolute magnitudes of galaxies: from --17 to --22 mag. An essential part of our sample (27--50\%) exhibit distorted spiral patterns. The galaxies do not show significant asymmetry in numbers of the ``S''- and ``Z''-like spin orientation. The mean  (pseudo)bulge-to-total mass ratio for the sample is estimated as 0.11. Due to a negligible internal extinction, low-light background, and small projection effect, the face-on Sc--Sd discs are suitable objects to recognize their central nuclei as moderate-mass BH candidates. About 40--60\% of the galaxies have distinct  unresolved nuclei, and their presence steeply depend on the luminosity of the host galaxy. 
\end{abstract}

\begin{keywords}
galaxies: bulges -- galaxies: spiral -- galaxies: nuclei 
\end{keywords}

\section{Introduction} The Hubble's morphological classification of galaxies and the improved classification by de Vaucouleurs et al. (1976) are based on the ratio of luminosities of spheroidal and disc components of a galaxy (the bulge-to-disc ratio). This ratio can be easily estimated visually in the case of ``edge-on'' galaxies. Karachentsev et al. (1993, 1999) compiled catalogues classifying such galaxies. The RFGC catalogue contains 4236 galaxies with angular diameters of more than $0.6^{\prime}$ and the apparent axial ratios $a/b>7$ distributed over the whole sky except for the Milky Way region. Later, Karachentseva et al. (2016)  compiled a sample of 817 extremely thin spiral galaxies with the axial ratios $(a/b)_B>10.0$ and $(a/b)_R>8.5$ in the blue and red bands of the Palomar Sky Survey, respectively. Over 80\% of such ultra-flat (UF) galaxies refer to the following morphological types: Sc (T=5), Scd (T=6), and Sd (T=7). These objects are interesting for almost full absence of a spheroidal stellar subsystem in them. Occurrence of a large number of such bulgeless galaxies impose  challenge to hierarchic clustering models of galaxy formation (Kormendy et al. 2010). The UF galaxies are isolated objects located in the low-density regions which makes it difficult to estimate their total mass  from orbital motions of rare satellites (Karachentsev et al. 2016). Studies of star-formation rates of UF galaxies are complicated by the presence of a considerable internal light extinction, especially in the far ultraviolet (Melnyk et al. 2017). 

The projection effect and internal extinction conceal important structural elements of thin edge-on galaxies: central unresolved nuclei, bars, and specific features of the spiral pattern. In order to estimate the statistical abundance of those structural elements of UF galaxies, a quite representative sample of face-on bulgeless galaxies is required. The present paper considers such a sample. The appearance of the following high angular resolution sky surveys in various optical bands facilitated its creation: SDSS (Abazajian et al. 2009) and Pan-STARRS1 (Chambers et al. 2016). Going forward, we intend to use the sample of face-on bulgeless galaxies to derive a more accurate estimate of the internal extinction in galactic discs, to estimate the characteristic star-formation rate in them, as well as to establish a membership of bulgeless galaxies in various elements of the cosmic large-scale  structure. 

\begin{figure}
\includegraphics[scale=4.0]{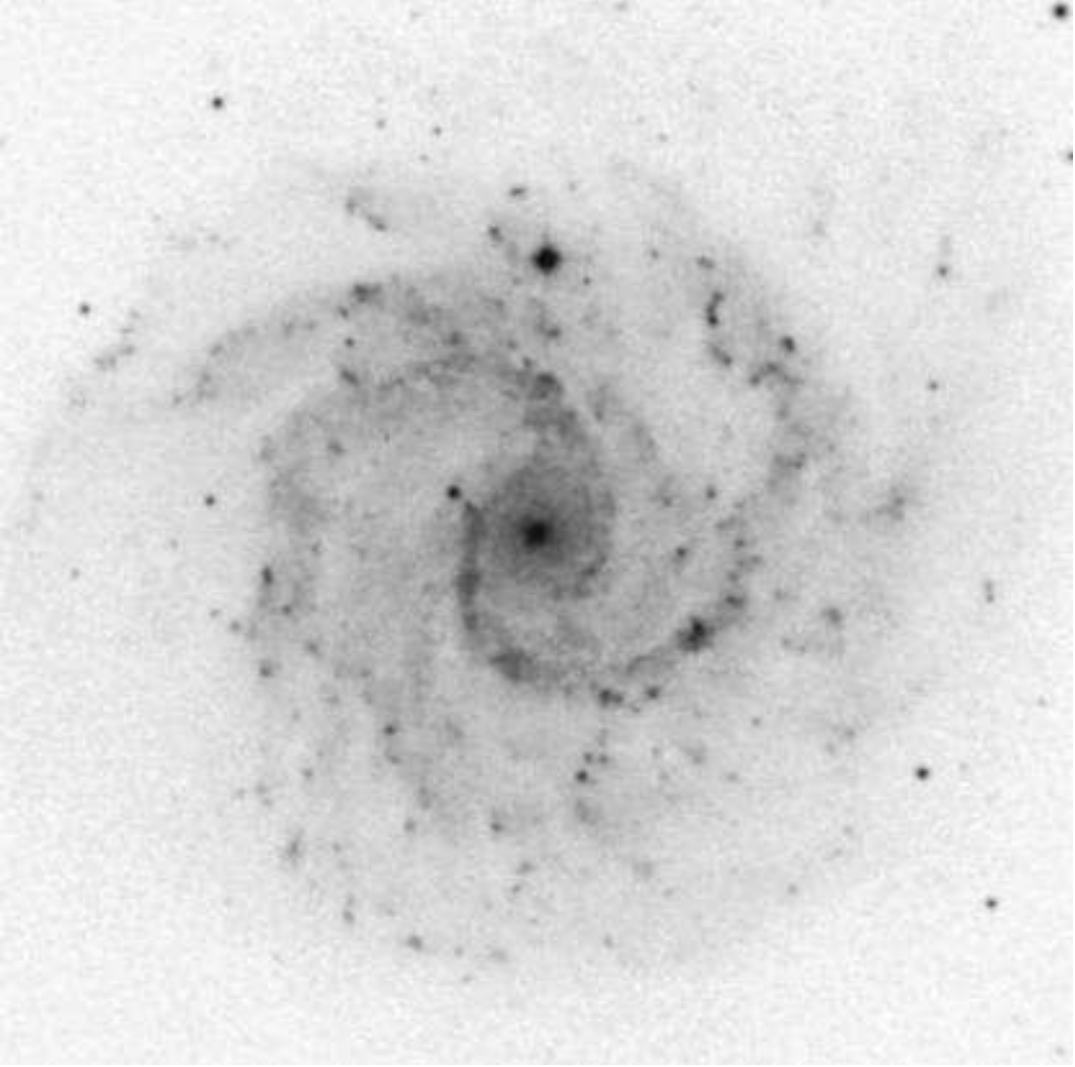} \caption{Face-on Scd galaxy NGC~3184. Reproduction from POSS-II of the $7^{\prime}\times7^{\prime}$ size, North -- at the top, East -- on the left.} \end{figure}

\section{Sample selection} To compile an analogue of the sample of UF galaxies, not edge-on, but face-on, we have undertaken the following procedures. We selected the galaxies in the HyperLEDA database (Makarov et al. 2014), which meet the conditions: \begin{equation} \log(r_{25})\leq0.05, \end{equation} 

where $r_{25}=a/b$ is the apparent axial ratio (major/minor axis) measured at a standard isophote level of $25^m/\sq^{\arcsec}$ in the B band; \begin{equation} \log(d_{25})\geq0.90, \end{equation} $d_{25}$ is the standard angular diameter (major axis) in units $0.1^{\prime}$ and determined from the same isophote; \begin{equation} 4.0<T<8.5, \end{equation} i.e., the morphological type according to the de Vaucouleurs's scale covers the range from Sbc to Sdm/Sm; \begin{equation} V_{LG}<10000\, {\rm km/s}, \end{equation} $V_{LG}$ is the radial velocity of a galaxy relative to the Local Group centroid. 

Our result was the sample of 537 objects. Further, we examined visually all the selected galaxies and left only the cases with the refined morphological types \begin{equation} 5\leq T\leq 7. \end{equation} 

As follows from the data (Heidmann et al. 1972, Karachentsev et al. 2017), the Sc, Scd, and Sd galaxies have the highest true axial ratio $a/b$ and, consequently, the minimum bulges. 

For reasons of homogeneity, in the final sample we left only the objects with the declination \begin{equation} DEC>-30^{\circ}, \end{equation} which are given in the Pan-STARRS1 survey that we used to classify the structural features of spiral galaxies. It resulted in decreasing our sample to 220 galaxies. Table 1 presents the list of these Sc--Scd--Sd galaxies viewed at a slight angle. The list includes seven bright nearby galaxies: IC~342 (3.28 Mpc), NGC~5068 (5.15 Mpc), M~101 (6.95 Mpc), NGC~6946 (7.73 Mpc), NGC~3344 (9.82 Mpc), NGC~628 (10.19 Mpc), and NGC~3184 (11.12 Mpc), the distances to which are measured with an accuracy of $\sim$5\% from the tip of the red giant branch, cepheids or supernovae. Figure 1 which is the reproduction from the Palomar Sky Survey (POSS-II) shows the image of the last one.

Table 1 presents the observed main characteristics of 220 face-on galaxies described in the Table Notes.

{
\small
 \begin{longtable}{lcrcclccccc}\\
 \caption{Basic properties of the  face-on bulgless galaxies.} \\ \hline
Name	    &$B_t$ 	&$V_{LG}$&	$V_{max}$& modD	&T& Bar  &Spin &Asm	&Nuc   &$M_B$\\
 \hline
 & mag   &km/s   &km/s&mag &    &     &   &     &      &mag\\
\hline
(1)			&(2)	&(3)    &(4)&(5)&  (6)& (7)&(8)&   (9)& (10)&	(11)\\	
\hline
\endfirsthead

\hline
(1)			&(2)	&(3)    &(4)&(5)&  (6)& (7)&(8)&   (9)& (10)&	(11)\\
\hline
\endhead 

\hline
 & & & & & & & & & & \\
\hline
\endlastfoot
NGC7816 	&13.91	&5443	&75	&34.42	&c	&0	&S  	&0	&2,2	&$-$20.87\\
UGC00044	&16.70	&6475	&40	&34.81	&d	&2	&S  	&0	&0,0	&$-$18.33\\
UGC00048	&15.60	&4596	&34	&34.07	&cd	&2	&Z	    &2	&0,0	&$-$19.15\\
NGC7834	    &15.35 	&5434	&52	&34.40	&cd	&1	&S  	&1	&1,0	&$-$19.76\\
NGC0039	    &14.34 	&5121	&80	&34.30	&c	&0	&Z	    &0	&2,2	&$-$20.38\\
UGC00160	&16.50	&4939	&55	&34.23	&cd	&0	&S  	&0	&1,0	&$-$17.97\\
PGC1075005	&15.57	&3378	&	&33.33	&cd	&0	&Z	    &1	&1,0	&$-$18.00\\
PGC002257	&15.41	&4290	&	&33.94	&c	&2	&Z	    &1	&0,0	&$-$18.64\\
IC1562	    &13.60 	&3771	&22	&33.56	&c	&0	&Z	    &1	&2,2	&$-$20.10\\
NGC0198	    &13.10 	&5414	&97	&34.40	&c	&0	&S  	&0	&2,2	&$-$21.48\\
IC0043	    &13.95 	&5103	&78	&34.28	&c	&2	&Z	    &1	&2,2	&$-$20.69\\
NGC0236	    &14.43	&5804	&52	&34.55	&c	&2	&S  	&2	&2,2	&$-$20.32\\
NGC0255	    &12.41	&1694	&68	&31.70	&c	&2	&Z    	&2	&1,1	&$-$19.49\\
IC0056	    &15.15	&6174	&60	&34.69	&cd	&0	&S  	&2	&2,1	&$-$19.75\\
UGC00626	&14.97	&5874	&	&34.58	&c	&0	&Z	    &0	&2,1	&$-$19.90\\
ESO412-013	&15.00	&5677	&	&34.51	&c	&0	&S  	&1	&1,1	&$-$19.67\\
ESO542-004	&15.00	&5657	&	&34.50	&d	&2	&S  	&2	&1,0	&$-$19.60\\
IC1666	    &14.35	&5108	&70	&34.30	&cd	&1	&S  	&0	&2,2	&$-$20.27\\
PGC005023	&15.78	&2372	&18	&32.51	&cd	&2	&S  	&0	&0,0	&$-$16.87\\
UGC00929	&14.82	&7581	&31	&35.15	&cd	&0	&Z	    &1	&2,2	&$-$20.60\\
NGC0575	    &13.72	&3337	&62	&33.33	&c	&2	&S  	&0	&2,2	&$-$19.94\\
UGC01087	&14.83	&4654	&59	&34.19	&c	&0	&S  	&1	&2,1	&$-$19.66\\
NGC0628	     &9.71	&828	&21	&30.03	&c	&0	&Z	    &0	&2,2	&$-$20.68\\
UGC01148	&15.45	&4951	&66	&34.30	&c	&0	&Z	    &1	&1,0	&$-$19.59\\
ESO543-021	&14.95	&5754	&	&34.54	&c	&1	&Z	    &1	&1,0	&$-$19.72\\
UGC01347	&13.50	&5759	&52	&34.57	&c	&2	&S	    &0	&2,2	&$-$21.44\\
PGC007210	&14.97	&8166	&99	&35.32	&c	&2	&S  	&2	&1,1	&$-$20.54\\
UGC01478	&14.56	&5024	&41	&34.24	&c	&2	&S  	&0	&1,1	&$-$20.25\\
UGC01546	&14.76	&2527	&38	&32.70	&c	&2	&S  	&0	&1,0	&$-$18.35\\
PGC007942	&15.00   &5332	&53	&34.37	&cd	&0	&S  	&0	&1,1	&$-$19.57\\
PGC008142	&14.91	&8061	&	&35.30	&c	&0	&Z	  &0	&1,1	&$-$20.61\\
UGC01626	&14.42	&5762	&73	&34.57	&c	&2	&S  	&1	&1,1	&$-$20.54\\
UGC02043	&14.96	&5316	&	&34.40	&c	&0	&Z	  &0	&2,2	&$-$20.05\\
IC0239	    &11.89	&1096	&48	&30.87	&cd	&2	&S  	&0	&1,0	&$-$19.34\\
UGC02094	&13.74	&5314	&116&34.39	&c	&0	&S  	&1	&2,2	&$-$20.98\\
UGC02121	&14.95	&6580	&36	&34.85	&c	&0	&S  	&0	&2,2	&$-$20.16\\
UGC02122	&14.41	&5243	&59	&34.35	&c	&0	&S  	&1	&2,2	&$-$20.66\\
ESO479-022	&15.39	&7266	&	&35.07	&c	&2	&S  	&1	&0,0	&$-$19.89\\
UGC02174	&15.09	&5296	&39	&34.39	&c	&1	&S  	&0	&2,1	&$-$19.92\\
ESO546-011	&14.52	&4535	&27	&34.01	&c	&2	&Z	    &1	&1,0	&$-$19.74\\
NGC1067	    &14.55	&4700	&59	&34.11	&c	&2	&Z	  &0	&2,2	&$-$20.43\\
UGC02323	&15.66	&8139	&59	&35.33	&c	&2	&Z	  &0	&2,2	&$-$20.84\\
UGC02399	&15.45	&8065	&86	&35.31	&c	&2	&S  	&1	&2,2	&$-$20.79\\
UGC02434	&16.00  &8108	&87	&35.32	&cd	&2	&S  	&1	&1,1	&$-$20.21\\
PGC012008	&15.00  &9439	&	&35.66	&c	&1	&S  	&1	&2,2	&$-$20.95\\
UGC02623	&15.59 	&4577	&46	&34.07	&d	&2	&S  	&1	&0,1	&$-$19.57\\
UGC02671	&17.00  &7205	&79	&35.07	&d	&1	&Z	    &1	&1,1	&$-$19.89\\
UGC02692	&14.07 	&6337	&82	&34.77	&c	&0	&Z	    &1	&2,2	&$-$21.10\\
UGC02712	&15.50  &7140	&69	&35.04	&cd	&2	&S  	&1	&2,1	&$-$20.82\\
UGC02721	&16.40	&6506	&25	&34.84	&d	&2	&Z	    &2	&0,0	&$-$19.12\\
NGC1325A	&13.40	&1262	&18	&31.11	&d	&0	&S	    &1	&2,0	&$-$17.87\\
ESO548-035	&13.49	&4161	&56	&33.82	&c	&2	&Z	    &1	&2,2	&$-$20.67\\
NGC1376	    &12.85	&4137	&70	&33.82	&c	&0	&S	    &2	&2,2	&$-$21.21\\
PGC013714	&15.10	&4308	&	&33.91	&cd	&0	&Z	    &1	&2,1	&$-$19.21\\
IC0342	     &9.68	 &239	&72	&27.68	&c	&0	&S	    &1	&2,2	&$-$20.45\\
UGC02859	&16.00  &5645	&88	&34.54	&c	&2	&Z	  &0	&2,2	&$-$20.51\\
UGC02932	&15.11	&4995	&88	&34.26	&d	&0	&S	  &0	&1,0	&$-$19.92\\
IC0370	    &14.94	&9649	&57	&35.72	&c	&2	&Z	  &0	&2,2	&$-$21.33\\
UGC03051	&16.00  &6900	&87	&34.98	&cd	&0	&Z	    &1	&1,1	&$-$20.00\\
NGC1599	    &14.10	&3947	&41	&33.73	&c	&0	&S	    &2	&2,1	&$-$19.86\\
PGC015609	&15.37	&9253	&	&35.63	&c	&2	&S	    &1	&2,2	&$-$20.63\\
IC2102	    &14.52	&3543	&34	&33.51	&cd	&2	&Z	    &2	&0,0	&$-$19.26\\
IC0391	    &12.98	&1768&	55	&32.12	&c	&0	&S	    &1	&1,1	&$-$19.73\\
ESO552-047	&15.26	&6568	&	&34.87	&c	&2	&Z	  &0	&2,1	&$-$20.12\\
UGC03308	&16.50	&8459	&45	&35.44	&d	&0	&S	  &0	&2,2	&$-$20.36\\
PGC017323	&13.49	&2016	&33	&32.30	&d	&0	&S	    &2	&1,0	&$-$19.20\\
PGC018031	&14.84	&6986	&71	&35.03	&c	&2	&S	    &1	&2,2	&$-$21.34\\
UGC03364	&16.00  &4630	&59	&34.15	&cd	&0	&S	  &0	&2,1	&$-$18.76\\
IC0441	    &14.40	&2042	&47	&32.34	&c	&2	&Z	  &0	&1,0	&$-$19.18\\
UGC03574	&13.20	&1551	&62	&31.34	&cd	&0	&S	    &1	&2,1	&$-$18.41\\
UGC03602	&14.76	&1983	&49	&32.34	&d	&0	&S	  &0	&0,0	&$-$18.17\\
UGC03703	&15.35	&7260	&22	&35.14	&c	&2	&S	  &0	&2,1	&$-$20.16\\
UGC03701	&14.80	&3085	&54	&33.30	&cd	&2	&Z	    &2	&1,0	&$-$18.78\\
UGC03806	&15.26	&5414	&90	&34.51	&cd	&2	&Z	    &1	&2,2	&$-$19.73\\
UGC03825	&15.06	&8303	&92	&35.43	&c	&2	&Z	    &1	&2,2	&$-$20.82\\
UGC03875	&15.24	&5303	&85	&34.46	&cd	&2	&S	    &1	&0,0	&$-$19.59\\
UGC03886	&16.00  &5006	&48	&34.33	&c	&2	&S	    &1	&2,0	&$-$18.66\\
UGC03924	&15.66	&5032	&41	&34.35	&d	&0	&S	    &1	&0,0	&$-$18.79\\
UGC04074	&13.75	&7263	&	&35.14	&c	&0	&S	    &1	&2,2	&$-$21.60\\
NGC2500	    &12.22	&558   & 41	&30.47	&d	&1	&S	    &1	&2,1	&$-$18.47\\
NGC2514	    &14.01	&4728	&50	&34.23	&c	&2	&Z	    &2	&2,1	&$-$20.43\\
UGC04107	&13.94	&3557	&63	&33.61	&c	&0	&Z	    &1	&2,2	&$-$19.92\\
PGC086610	&16.40	&4718	&21	&34.23	&d	&0	&Z	    &1	&0,0	&$-$18.05\\
PGC023378	&14.49	&4359	&50	&34.05	&cd	&0	&Z	    &1	&2,1	&$-$19.83\\
UGC04380	&15.05	&7554	&52	&35.24	&c	&0	&Z	  &0	&2,1	&$-$20.53\\
UGC04445	&15.06	&6435	&18	&34.88	&c	&0	&S	  &0	&2,1	&$-$20.19\\
IC0509	    &13.83	&5398	&36	&34.52	&c	&2	&Z	    &1	&2,1	&$-$20.97\\
NGC2607	    &14.95	&3447	&25	&33.56	&c	&0	&S	    &2	&2,2	&$-$18.86\\
UGC04536	&15.42	&7507	&	&35.22	&c	&2	&S	    &1	&2,2	&$-$20.31\\
NGC2661	    &13.86	&3958	&41	&33.87	&cd	&0	&Z	    &2	&2,2	&$-$20.23\\
UGC04669	&15.46	&3988	&68	&33.89	&d	&2	&Z	    &1	&0,0	&$-$18.63\\
UGC04853	&15.24	&2512	&28	&32.94	&cd	&2	&S	    &2	&0,0	&$-$17.99\\
PGC026687	&14.72	&3380	&	&33.54	&cd	&0	&Z	    &1	&1,0	&$-$19.04\\
UGC05015	&15.30	&1598	&51	&32.02	&cd	&0	&S	    &1	&1,0	&$-$16.85\\
UGC05153	&15.65	&8044	&	&35.38	&c	&0	&S	  &0	&2,1	&$-$19.95\\
UGC05169	&15.44	&7699	&	&35.29	&d	&2	&Z	  &0	&2,0	&$-$19.99\\
NGC2967  	&12.28	&1679	&57	&32.16	&c	&0	&S	    &1	&2,2	&$-$20.37\\
UGC05274	&14.92	&5764	&59	&34.68	&cd	&0	&Z	    &2	&2,1	&$-$19.99\\
ESO566-019	&13.99	&3429	&71	&33.56	&cd	&2	&Z	    &2	&1,0	&$-$19.83\\
ESO499-011	&15.01	&2342	&32	&32.77	&d	&2	&S	    &2	&0,0	&$-$18.07\\
PGC028556	&14.84	&7163	&26	&35.14	&c	&0	&Z	    &1	&2,1	&$-$20.62\\
ESO567-010	&14.44	&2779	&19	&33.13	&cd	&2	&Z	    &1	&1,0	&$-$19.05\\
PGC029301	&15.24	&9103	&	&35.66	&c	&2	&Z	  &0	&2,2	&$-$20.71\\
UGC05474	&14.89	&5847	&64	&34.71	&cd	&2	&S	    &1	&0,1	&$-$19.98\\
UGC05483	&15.13	&6014	&36	&34.77	&c	&0	&Z	  &0	&2,0	&$-$19.93\\
PGC029882	&14.54	&9046	&	&35.64	&c	&2	&S	  &0	&2,2	&$-$21.45\\
PGC029929	&14.30	&3214	&38	&33.44	&d	&2	&S	    &1	&2,0	&$-$19.56\\
NGC3184	    &10.41	 &588	&51	&30.33	&cd	&0	&Z	    &1	&2,2	&$-$20.01\\
PGC030452	&14.78	&6363	&	&34.90	&c	&0	&S	    &1	&1,1	&$-$20.27\\
PGC030830	&14.91	&7209	&	&35.14	&c	&0	&Z	    &1	&2,1	&$-$20.40\\
NGC3344	    &10.50	&499	&69	&29.96	&c	&2	&Z	  &0	&2,2	&$-$19.64\\
PGC031979	&14.05	&1838	&68	&32.34	&d	&0	&S	   & 2	&0,0	&$-$18.50\\
PGC032091	&14.45	&2268	&59	&32.77	&cd	&2	&S   	&2	&1,0	&$-$18.56\\
NGC3433  	&13.30	&2553	&116&33.01	&c	&0	&S   	&1	&2,2	&$-$19.89\\
PGC032817	&15.28	&8017	&	&35.39	&cd &2	&S  	&2	&2,1	&$-$20.37\\
NGC3506	    &13.16	&6252	&85	&34.87	&c	&0	&Z  	&2	&2,2	&$-$21.92\\
UGC06130	&15.02	&8151	&98	&35.42	&c	&2	&S  	&1	&2,0	&$-$20.62\\
UGC06194	&14.62	&2551	&48	&33.01	&cd	&2	&S  	&2	&0,0	&$-$18.49\\
PGC034006	&14.13	&7543	&73	&35.26	&c	&2	&S  	&1	&2,2	&$-$21.49\\
NGC3596	    &11.79	&1062	&50	&31.32	&cd	&0	&S  	&1	&2,2	&$-$19.68\\
UGC06326	&15.76	&2094	&53	&32.62	&d	&2	&Z  	&1	&0,0	&$-$17.00\\
UGC06335	&14.91	&3026	&23	&33.30	&cd	&0	&S   	&1	&1,0	&$-$18.49\\
UGC06429	&13.78	&3849	&23	&33.80	&c	&0	&Z  	&0	&2,1	&$-$20.14\\
IC0696	    &14.50	&6132	&74	&34.83	&cd	&2	&S  	&1	&1,1	&$-$20.60\\
UGC06528	&14.12	&3365	&15	&33.54	&c	&2	&S  	&1	&1,0	&$-$19.55\\
NGC3763	    &12.90	&5651	&100&34.65	&c	&2	&S  	&1	&2,2	&$-$21.99\\
PGC036269	&15.59	&6462	&86	&34.92	&c	&0	&S  	&1	&1,0	&$-$19.53\\
PGC036353	&14.14	&2868	&	&33.25	&c	&0	&S  	&1	&1,0	&$-$19.52\\
NGC3938	    &10.87	&841  	&38	&30.87	&c	&0	&Z  	&1	&2,2	&$-$20.12\\
ESO573-002	&15.50	&5550	&57	&34.06	&c	&0	&S  	&1	&2,1	&$-$19.49\\
NGC4136	    &11.90	&560 	&38	&31.09	&c	&1	&Z  	&2	&2,2	&$-$19.32\\
NGC4195	    &15.29	&4473	&33	&34.13	&cd	&2	&S  	&2	&0,0	&$-$19.06\\
NGC4303	    &10.16	&1422	&66	&31.36	&c	&2	&Z  	&1	&2,2	&$-$21.34\\
NGC4303A	&13.53	&1135	&50	&31.45	&cd	&0	&S  	&1	&0,0	&$-$18.07\\
IC3267	    &14.12	&1103	&24	&32.86	&c	&0	&S  	&1	&2,1	&$-$18.87\\
IC3271  	&14.57	&7083	&69	&35.13	&c	&2	&S  	&2	&1,1	&$-$20.79\\
NGC 4411B	&12.98	&1149	&35	&32.14	&cd	&0	&Z  	&1	&2,0	&$-$19.38\\
NGC4535	    &10.56	&1841	&122&31.05	&c	&2	&S  	&1	&2,2	&$-$20.65\\
NGC4571  	&11.92	&244	&67	&30.93	&cd	&0	&Z  	&2	&2,2	&$-$19.27\\
NGC4653	    &12.77	&2471	&92	&33.08	&c	&0	&Z  	&1	&2,2	&$-$20.50\\
PGC042868	&13.03	&1259	&49	&31.64	&d	&0	&S  	&2	&2,0	&$-$18.81\\
ESO574-029	&13.67	&6081	&	&34.80	&c	&1	&S  	&0	&2,2	&$-$21.45\\
NGC4688	    &12.92	&857  	&20	&31.03	&d	&2	&Z  	&2	&1,0	&$-$18.32\\
NGC4900	    &11.89	&836 	&36	&31.86	&c	&2	&S  	&1	&1,2	&$-$20.11\\
UGC08153	&14.49	&2742	&54	&33.16	&cd	&2	&Z  	&2	&0,0	&$-$18.81\\
PGC045690	&15.86	&4892	&	&34.33	&c	&0	&Z  	&0	&1,0	&$-$18.59\\
NGC5068	    &10.64	&471	&38	&28.56	&cd	&2	&S  	&2	&0,1	&$-$18.45\\
UGC08436	&15.03	&3019	&28	&33.30	&d	&0	&Z  	&0	&0,0	&$-$18.43\\
NGC5154	    &14.73	&5586	&64	&34.62	&c	&0	&S  	&1	&2,2	&$-$20.03\\
PGC048087	&14.45	&2315	&40	&32.81	&d	&2	&Z  	&1	&1,0	&$-$18.75\\
NGC5260	    &13.58	&6313	&119&34.87	&c	&2	&Z  	&0	&2,2	&$-$21.69\\
ESO445-76	&14.73	&2511	&44	&32.94	&d	&0	&S  	&1	&0,0	&$-$18.57\\
UGC08877	&15.18	&2464	&17	&32.91	&d	&2	&S  	&0	&0,0	&$-$17.84\\
NGC5405	    &14.52	&6846	&43	&35.06	&c	&0	&Z  	&1	&2,2	&$-$20.78\\
PGC049982	&16.00  &9208	&	&35.67	&c	&0	&S  	&0	&2,0	&$-$19.52\\
NGC5457  	&8.36   &374	&75	&29.26	&c	&1	&S  	&2	&2,2	&$-$20.97\\
NGC5434	    &13.94	&4587	&18	&34.21	&c	&0	&Z  	&0	&2,1	&$-$20.46\\
UGC09008	&15.29	&5305	&55	&34.52	&cd	&0	&S  	&0	&1,0	&$-$19.35\\
NGC5468 	&12.95	&2734	&54	&33.40	&c	&1	&Z  	&2	&2,2	&$-$20.63\\
NGC5476	    &13.34	&2530	&80	&32.97	&cd	&0	&Z  	&1	&1,2	&$-$19.81\\
NGC5494 	&13.30	&2428	&104&32.99	&c	&0	&S  	&1	&2,2	&$-$20.08\\
ESO446-031	&13.60	&2471	&51	&32.91	&cd	&2	&Z  	&1	&1,0	&$-$19.69\\
UGC09144	&15.95	&7883	&	&35.35	&cd	&0	&S  	&1	&2,0	&$-$19.50\\
UGC09216	&14.51	&5618	&106&34.63	&c	&0	&S  	&0	&2,0	&$-$20.25\\
NGC5660 	&12.38	&2458	&55	&32.91	&c	&0	&Z  	&1	&2,2	&$-$20.70\\
UGC09317	&15.00	&4476	&21	&34.15	&cd	&2	&S  	&1	&0,0	&$-$19.33\\
ESO580-014	&14.55	&5955	&37	&34.74	&cd	&2	&S  	&1	&1,2	&$-$20.67\\
UGC09837	&13.81	&2849	&71	&33.19	&c	&2	&S  	&2	&1,1	&$-$19.52\\
UGC09945	&14.19	&6857	&97	&35.04	&c	&0	&Z  	&1	&2,2	&$-$21.16\\
IC1132  	&14.37	&4604	&45	&34.19	&c	&2	&Z  	&1	&2,1	&$-$20.17\\
UGC09982	&15.26	&7896	&45	&35.33	&cd	&2	&Z  	&0	&1,0	&$-$20.21\\
NGC5989 	&13.56	&3079	&68	&33.33	&cd	&0	&S  	&1	&2,1	&$-$19.92\\
UGC10020	&14.44	&2175	&25	&32.66	&d	&0	&S  	&1	&1,0	&$-$18.51\\
PGC056010	&15.42	&4619	&30	&34.21	&cd	&2	&S  	&0	&0,0	&$-$19.01\\
PGC056318	&15.33	&5998	&	&34.75	&c	&2	&S  	&0	&2,0	&$-$19.60\\
PGC056639	&15.77	&8371	&	&35.47	&cd	&2	&S  	&0	&2,2	&$-$20.78\\
NGC6143 	&13.92	&5521	&124&34.57	&c	&2	&S  	&1	&2,2	&$-$20.80\\
PGC058201	&15.69	&8562	&54	&35.51	&c	&2	&Z  	&1	&1,2	&$-$20.12\\
UGC10427	&15.19	&9036	&57	&35.63	&c	&2	&S  	&2	&2,2	&$-$20.61\\
IC1221  	&14.59	&5706	&18	&34.63	&cd	&0	&S  	&1	&2,2	&$-$20.21\\
UGC10590	&14.06	&3287	&	&33.46	&cd	&2	&Z  	&0	&1,1	&$-$19.60\\
IC1236  	&14.23	&6171	&45	&34.80	&c	&2	&Z  	&0	&2,2	&$-$21.01\\
AGC260883	&16.07	&9059	&29	&35.63	&cd	&0	&S  	&1	&0,0	&$-$19.73\\
UGC10956	&15.64	&6825	&	&35.00  &c	&2	&S  	&0	&2,2	&$-$20.61\\
NGC6493 	&15.44	&6199	&	&34.80	&cd	&2	&S  	&2	&2,1	&$-$19.58\\
UGC11064	&14.40	&7257	&101&35.05	&c	&2	&Z  	&0	&2,2	&$-$20.92\\
UGC11214	&15.00  &2834	&36	&33.10	&cd	&2	&Z  	&2	&0,0	&$-$19.15\\
UGC11302	&16.00  &4324	&55	&33.99	&d	&2	&S  	&1	&0,0	&$-$18.77\\
NGC6711 	&13.71 	&4944	&84	&34.30	&c	&2	&Z  	&1	&2,2	&$-$21.02\\
UGC11430	&14.41	&5769	&47	&34.62	&c	&0	&Z  	&2	&2,2	&$-$20.68\\
NGC6821 	&13.62	&1680	&54	&31.86	&d	&2	&Z  	&2	&0,0	&$-$19.70\\
NGC6946 	& 9.75	& 349	&98	&29.44	&c	&0	&Z  	&2	&2,2	&$-$21.20\\
PGC902799	&16.30	&3584	&42	&33.51	&c	&0	&S  	&1	&0,0	&$-$17.50\\
UGC11636	&16.00  &2859	&62	&33.10	&cd	&2	&S  	&1	&0,0	&$-$19.14\\
PGC065744	&14.62	&5810	&	&34.58	&c	&0	&Z  	&0	&2,2	&$-$20.38\\
PGC3083038	&14.44	&8486	&	&35.43	&c	&2	&Z  	&1	&2,2	&$-$21.39\\
NGC7137  	&13.05	&1977	&46	&32.16	&c	&2	&Z  	&0	&2,2	&$-$19.79\\
UGC11816	&14.83	&4962	&52	&34.26	&c	&2	&Z  	&1	&1,1	&$-$20.10\\
UGC11834	&14.45	&3554	&79	&33.49	&c	&2	&Z  	&0	&1,0	&$-$19.41\\
ESO532-008	&15.29	&6365	&	&34.78	&c	&2	&Z	    &1	&0,0	&$-$19.75\\
IC1418	    &15.23	&8552	&93	&35.44	&c	&0	&Z  	&2	&2,2	&$-$20.47\\
PGC068549	&14.23	&5163	&	&34.30	&c	&2	&S  	&2	&1,2	&$-$20.58\\
ESO602-027	&14.48	&5835	&21	&34.57	&c	&2	&S  	&2	&0,1	&$-$20.38\\
NGC7309	    &13.04	&4168	&49	&33.82	&c	&2	&S  	&2	&2,2	&$-$21.12\\
PGC133417	&14.97	&8965	&	&35.54	&c	&2	&S  	&1	&2,1	&$-$20.73\\
UGC12156	&14.76	&5483	&53	&34.47	&c	&2	&S  	&1	&2,1	&$-$20.34\\
ESO603-011	&15.01	&8303	&	&35.37	&c	&2	&Z  	&1	&2,1	&$-$20.64\\
UGC12192	&16.50	&6789	&25	&34.94	&d	&2	&Z  	&1	&0,0	&$-$19.05\\
IC5261   	&13.88	&3347	&65	&33.30	&cd	&2	&Z  	&2	&2,1	&$-$19.68\\
NGC7437 	&13.95	&2364	&44	&32.55	&cd	&2	&S  	&2	&2,0	&$-$18.81\\
NGC7495 	&13.76	&5133	&85	&34.28	&c	&2	&Z  	&1	&2,2	&$-$21.02\\
NGC7535 	&14.28	&4884	&50	&34.17	&cd	&2	&Z  	&2	&2,1	&$-$20.20\\
UGC12522	&15.41	&3033	&47	&33.10	&d	&2	&S  	&1	&0,0	&$-$18.21\\
UGC12585	&14.63	&3899	&40	&33.66	&d	&2	&S  	&1	&1,0	&$-$19.42\\
UGC12635	&15.31	&5400	&36	&34.40	&cd	&0	&S  	&0	&2,1	&$-$19.38\\
PGC071751	&14.74	&8381	&92	&35.40	&c	&0	&Z  	&2	&2,2	&$-$21.01\\
ESO605-016	&13.23	&7999	&100&35.28	&c	&2	&Z  	&1	&2,2	&$-$22.24\\
PGC072738	&14.82	&6024	&59	&34.65	&c	&2	&S  	&2	&2,2	&$-$20.15\\
UGC12838	&14.86	&7375	&56	&35.09	&c	&0	&S  	&1	&2,2	&$-$20.52\\
NGC7798 	&12.95	&2650	&34	&32.81	&c	&0	&S  	&1	&2,2	&$-$20.23\\ \hline
\end{longtable}}


 {\bf Notes:} Columns contain the following data. (1) -- the name of the galaxy in HyperLEDA. 
(2) -- the apparent integrated B magnitude from HyperLEDA with some refinements 
from NED (http://ned.ipac.caltech.edu), and other sources. (3) -- the radial 
velocity relative to the Local Group centroid. (4) -- the rotation amplitude of 
a galaxy from HyperLEDA. (5) -- the distance modulus from HyperLEDA; in some cases: 
NGC~2500, NGC~4136, NGC~4303, NGC~4535, NGC~4571, NGC~4653, NGC~4900, and NGC~6946, 
the distance moduli were updated with the latest observational data. (6) -- the 
morphological type: Sc, Scd, or Sd. (7) -- the presence (``2'') or absence (``0'') 
of a bar; uncertain cases of a diffuse or short bar are marked with ``1''. (8) -- 
the ``S''- or ``Z''-like spin orientation of spiral arms. (9) -- the presence 
(``2'') or absence (``0'') of a significant asymmetry of the spiral structure; 
(10) -- the presence (``2'') or apparent absence (``0'') of an unresolved nucleus 
in a galaxy, while ``1'' characterizes the central concentration of a diffuse 
shape. The first digit refers to visibility in the optical range in the Pan-STARRS1 
images, the second digit characterizes the visibility of the nucleus in the $K_s$ 
band according to 2MASS (Jarrett et al. 2000). Case ``2,2'' corresponds to the 
presence of a distinct unresolved nucleus (a black hole candidate) in all the 
Pan-STARRS1 and 2MASS photometric bands. (11) -- the absolute B band magnitude 
of a galaxy corrected for the Galactic extinction according to Schlegel et al. (1998). 
\begin{table} 
 \caption{Numbers of face-on Sc--Scd--Sd galaxies with different properties.}
 \begin{tabular}{llrr} \hline

Property &&N &(\%)\\ \hline

Type & Sc &127 &58\\ & Scd &62 &28\\ & Sd &31 &14\\ \hline

Bar & 0 &95 &43\\ & 1 &11 &5\\ & 2 &114 &52\\ \hline

Spin & S &123 &56\\ & Z &97 &44\\ \hline

Asymmetry & 0 &60 &27\\ & 1 &110 &50\\ & 2 &50 &23\\ \hline

B/Tot & 0.05 &15 &7\\ & 0.10 &124 &56\\ & 0.15 &64 &29\\ & 0.20 &17 &8\\ \hline

Nucl., Pan-STARRS1: & 0 &38 &17\\ & 1 &53 &24\\ & 2 &129 &59\\ 

Nucl., 2MASS& 0 &78 &35\\ & 1 &56 &26\\ & 2 &86 &39\\ \hline
\end{tabular} 
\end{table} 
\begin{table} 
\caption{Numbers of face-on bulgeless galaxies with different nuclei visibility in the Pan-STARRS1 and 2MASS surveys.}
\begin{tabular}{l|lrrr}\hline 
Pan-STARRS1: &2 &13 &35 &81 \\ 
&1 &31 &17 &5 \\ 
&0 &34 &4 &0 \\ \hline
\multicolumn{1}{l}{2MASS:}&
&\multicolumn{1}{r}{0}&
 \multicolumn{1}{r}{1}&
 \multicolumn{1}{r}{2}\\  
\hline
\end{tabular}
\end{table} 
\label{lastpage}
Table 2 presents the distribution of the galaxies from our sample by various structural features. As one can see, this sample demonstrates a considerable variety of structural forms of galaxies. 
\begin{figure*} 
\includegraphics[scale=0.55]{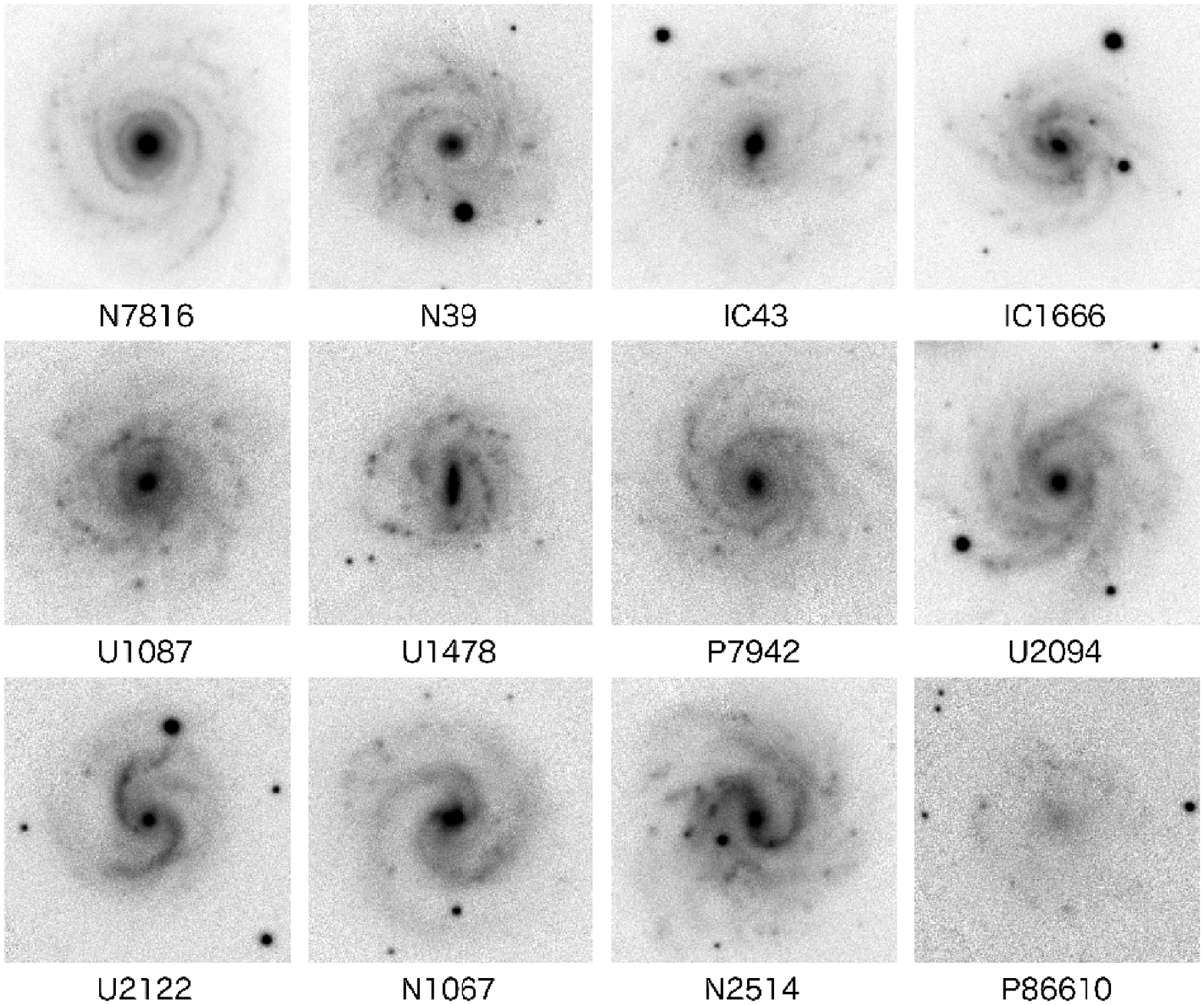}\\
\includegraphics[scale=0.55]{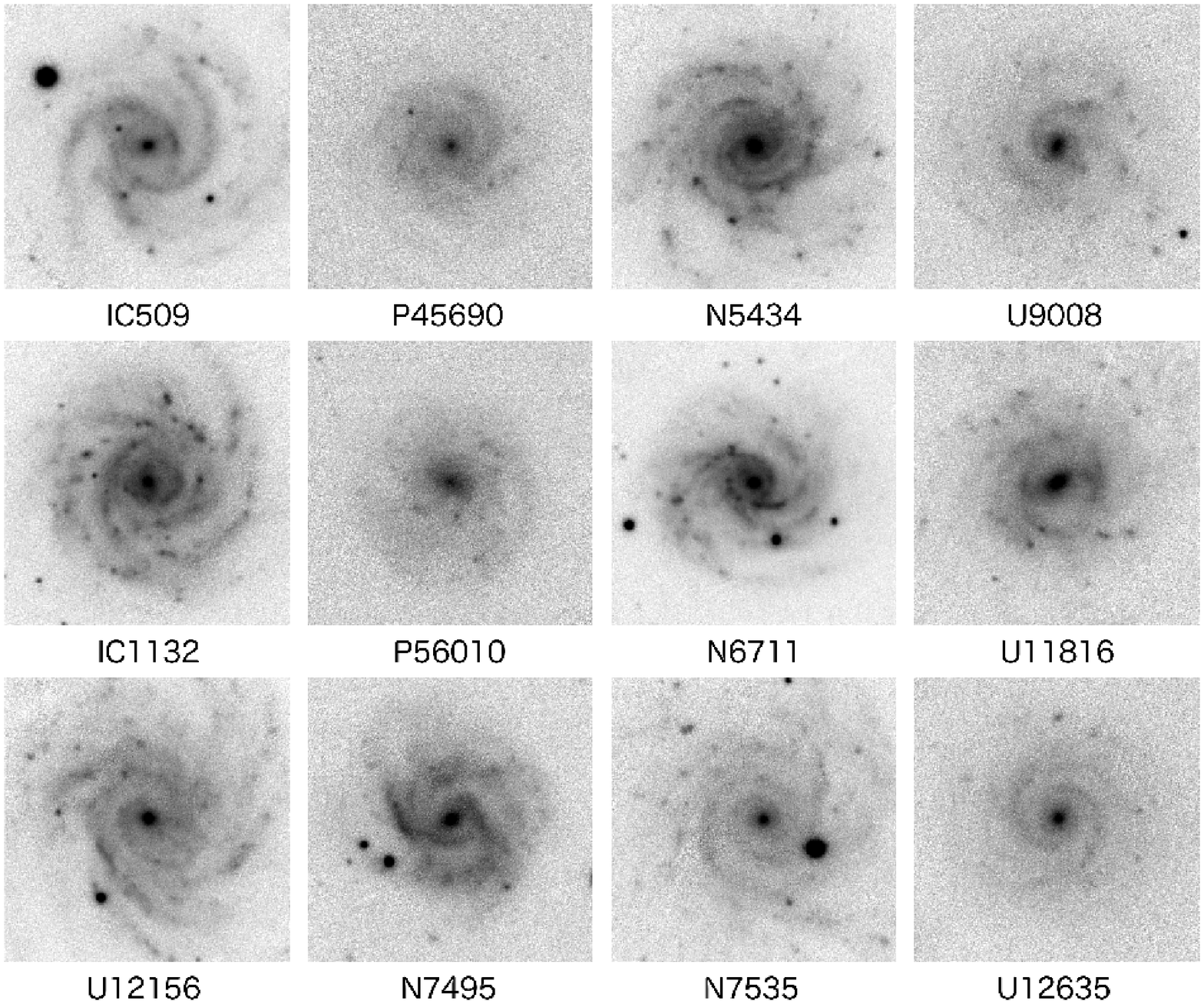}
\caption{Mosaic images of 24 face-on galaxies of the Sc--Scd--Sd types with radial velocities in the range of 4500-5500 km/s. The reproductions are made from the images in the $g$ band in the Pan-STARRS1 survey of the $75^{\prime\prime}\times75^{\prime\prime}$ size, North -- at the top, East -- on the left.} \end{figure*} 
Figure 2 shows a mosaic of images of 24 face-on galaxies.  The galaxies in the mosaic are selected by the radial velocities in the range of [4500--5500] km/s, in which the median velocity lies. Here we do not show the images having corrupted sky background in Pan-STARRS1. Galaxy images of the sizes of $75^{\prime\prime}\times 75^{\prime\prime}$ correspond to the $g$- band of the Pan-STARRS1 survey. In some galaxies, the spiral structure extends beyond the chosen format. 
\begin{figure} \includegraphics[scale=0.6]{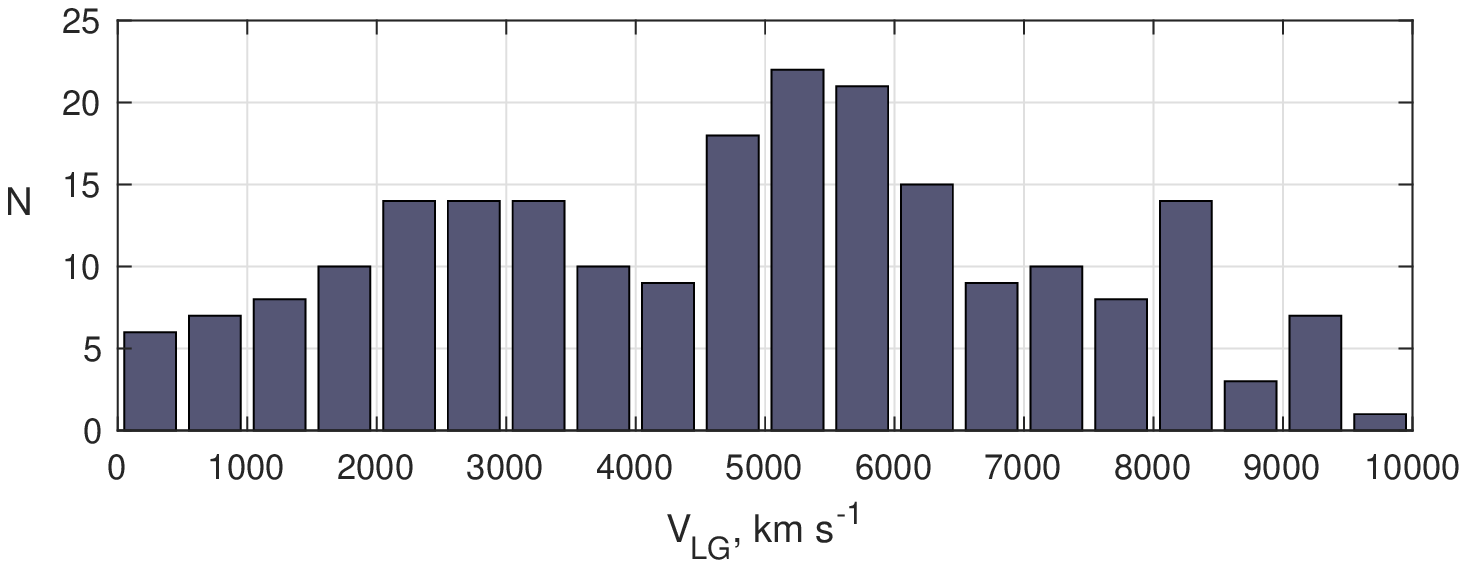} \includegraphics[scale=0.6]{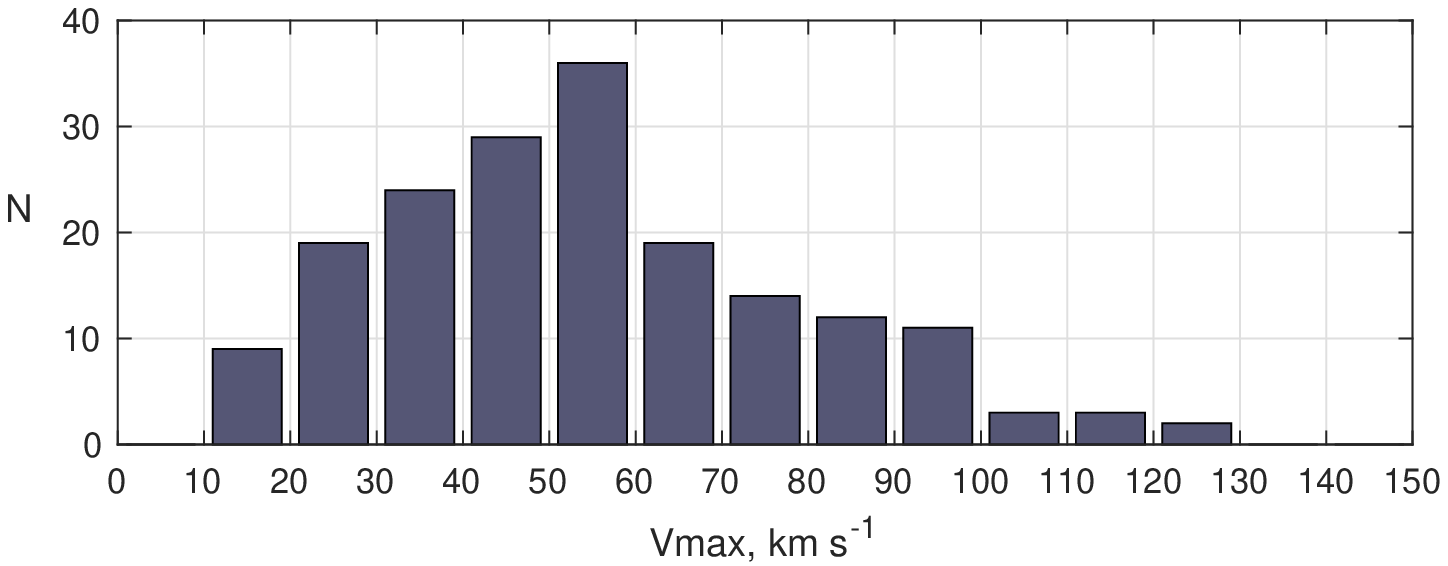} \caption{Distribution of face-on bulgeless galaxies by radial velocities (the top panel) and by rotation amplitudes (the bottom panel).} \end{figure} 

\section{Basic properties of the sample} The top panel of Fig.3 gives the distribution of the sample galaxies by radial velocities.  The mean value for them, $\langle V_{LG}\rangle =+4845\pm170$ km/s, within a standard error, coincides with the mean for the sample of 284 Sc, Scd, and Sd galaxies, $+5153\pm200$ km/s, classified as ultra-flat (Karachentseva et al. 2016). It follows that both samples are taken from volumes having approximately the same depth. Here we should notice that the UF galaxies were selected by the angular diameter, $a>1.2^{\prime}$, which is 1.5 times greater than the assumed minimum diameter of face-on galaxies. This difference was justified by the fact that the $90^{\circ}$-turn of a face-on galaxy exactly one and a half times increases its apparent major diameter. Thus, selecting the major diameter $(d_{25})_{min}=0.8^{\prime}$ under condition (2) provided a sufficient depth and representativeness of the face-on bulgeless galaxies sample. 

The bottom panel of Fig. 3 presents the distribution of our face-on galaxies by the observed rotation amplitude $V_m$. The pattern of this distribution with a mean of 56 km/s is determined not so much by masses of the galaxies but their small inclination angles. The mean full rotation amplitude of edge-on UF galaxies is $\langle V_m\rangle =142$ km/s from the data of Table 1 (Karachentseva et al. 2016). Assuming it as a typical face-on galaxy, we obtain an estimate of the mean inclination angle in our sample $i=\arcsin(56/142)=23^{\circ}$. With this characteristic inclination angle, the apparent axial ratio of the galaxy is $\log(r_{25})\simeq0.037$. Comparing it with condition (1), we can conclude that due to measurement errors of $r_{25}$ a part of galaxies with the true inclination angle $i\simeq0$ could fail to get into the sample under consideration. Also, omission of true face-on galaxies can result 
not only from errors in the axial ratio measurements but from the fact that many late-type galaxies are intrinsically lopsided. Then apparent axial ratios
do not tell us inclinations in the normal way for circular discs.

As follows from the data given in Table 2, the majority of face-on galaxies (58\%) were classified as the Sc-type spirals, while 28\% of them are of the Scd type, and 14\% -- of the Sd type. This proportion is quite similar to the proportion of types of edge-on galaxies in the RFGC catalogue: 49\% (Sc), 35\% (Scd), and 16\% (Sd), however, it differs markedly from the type proportion of UF galaxies: 25\% (Sc), 28\% (Scd), and 47\% (Sd). Comparing our classification with that in HyperLEDA shows that the mean-square error in determining the morphological type of a galaxy is $\sigma (T)=0.7$. 
\begin{figure} \includegraphics[scale=0.7]{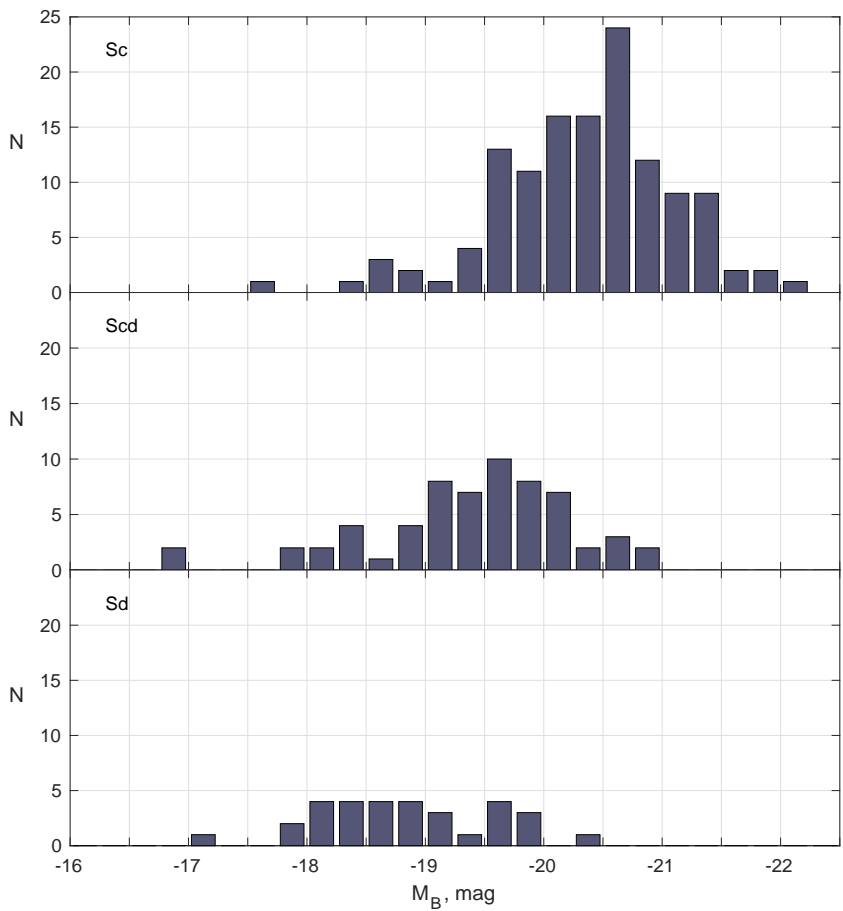} \caption{Distribution of face-on galaxies of the Sc, Scd, and Sd types by absolute magnitudes.} \end{figure} 
Three panels of Fig. 4 show the distribution of face-on bulgeless galaxies of different types by the absolute magnitude $M_B$.  The histograms show the  well-known decrease of the mean luminosity (Kormendy, 1979; Karachentsev et al. 2017) along the Hubble sequence: $\langle M_B\rangle=-20.37\pm0.07$ for Sc, $-19.40\pm0.10$ for Scd, and $-18.82\pm0.13$ for Sd galaxies. 

Favourable aspect angles and the absence of bright bulges make the face-on bulgeless galaxies suitable for statistical studies of the occurrence of bars in them. Except for a small number of cases (5\%), the presence or absence of a bar in our sample galaxies is clearly distinguishable. Almost half of the face-on bulgeless galaxies (43\%) demonstrate extended or short central bars. Figure 5 presents the distribution of galaxies with bars (the top panel) and without any (the bottom panel). As one can see, the bars are found both in high-luminosity galaxies and in dwarf discs. Both distributions are similar with a slight shift, $\Delta M_B=-0.15^m\pm0.13^m$, towards the objects with bars. 
\begin{figure} \includegraphics[scale=0.7]{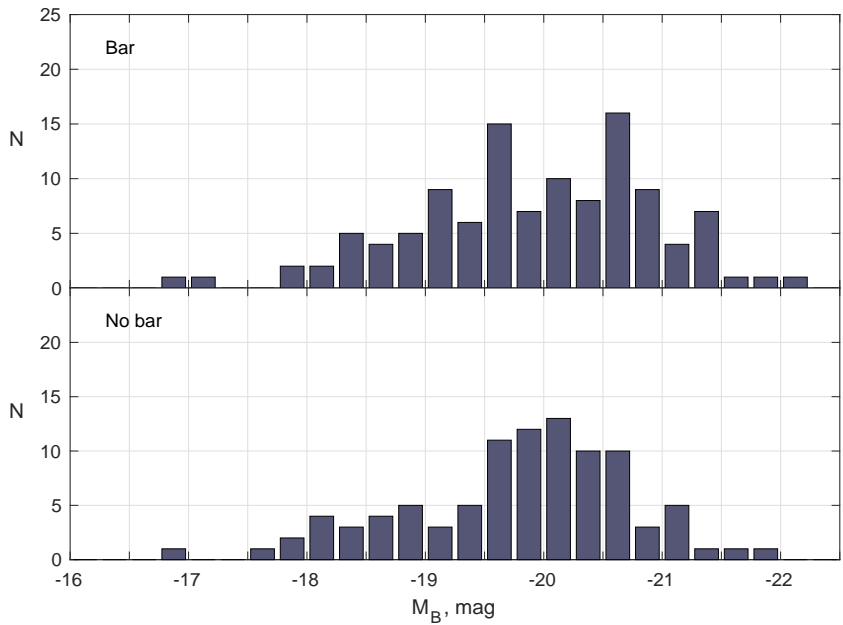} \caption{Distribution of face-on bulgeless galaxies with the signs of bar presence (the top panel) and without any (the bottom panel) by absolute magnitudes.} \end{figure} 
Most of the considered galaxies look like isolated objects. According to our preliminary estimations, they have no nearest significant neighbours within a radius of 100--200 kpc. Therefore, it can be affirmed that the presence of a bar in a thin disc is their intrinsic dynamic property almost independent of luminosity or linear sizes of the disc. Moreover, the fact that so many of these galaxies are isolated also suggests that their bars are not produced
as a result of tidal force.

In modern models of the formation of galactic discs with their large angular momentum, the possibility of an ordered orientation of the galaxy spins with respect to the large-scale structure elements (filaments and walls) is allowed. Such a space organization of spins could manifest itself in the form of an excessive number of galaxies with ``S''- or ``Z''-like structures. The data in Table 2 show that the number of galaxies with different configuration of spiral arms, $N_S:N_Z =123:97$, show only a statistically insignificant asymmetry on scales of $\sim5000$ km/s. However, the absence of a preferred spin orientation should be controlled with a more representative sample of face-on galaxies. Parnovsky et al. (1994) noticed some evidence of anisotropy in the distribution of planes of edge-on galaxies in the FGC catalogue.  

The spiral structure of our sample galaxies is characterized by varying degrees of asymmetry. About a quarter of galaxies (27\%) are of a symmetric shape. Galaxies from another part of the sample (23\%) show quite a strong asymmetry of spiral arms. The rest half of the sample galaxies have feebly-marked signs of asymmetry. The galaxies NGC 628 and NGC 3344 can be an example of a symmetrical spiral pattern. Considerable distortions of the spiral structure can be seen in the galaxy M 101 (NGC 5457); their apparent cause is the disturbance from the nearby companion NGC 5474, the structure of which is subject to even stronger tidal distortion. Statistics of absolute magnitudes show that the luminosity of strongly distorted galaxies, $\langle M_B\rangle = -19.75^m\pm0.14^m$, is slightly lower than that of objects of a symmetrical structure, $\langle M_B\rangle = -20.05^m\pm0.15^m$. 
     
\section{Bulge-to-disc ratio} Based on the results of the surface photometry of 98 galaxies, Simien \& de Vaucouleurs (1986) performed the decomposition of galaxies into the components: a bulge and a disc. According to their data, the luminosity ratio of bulge-to-disc in the blue band can be expressed as 

$$B/D=0.754\times 10^{-T/5}.$$ For late-type galaxies, this yields the ratio of the bulge luminosity to the total luminosity: 0.070 for Sc, 0.045 for Scd, and 0.029 for Sd discs. 
Subsequently, Oohama et al. (2009) decomposed 737 galaxies of the morphological types $T=1\div4$ using the SDSS photometric data in various bands. Extrapolation of their estimates of the B/D ratio towards later types with $T>4$ results in a bulge proportion in the total luminosity of less than 10\%. The similar work by Kim et al. (2016) on decomposition of 14233 SDSS galaxies in the $r$ band gave the average relative luminosity of bulges of Sc galaxies less than 20\%. 

 However, as it was shown by Kormendy (1982, 1993, 2013) and Kormendy
\& Kennicutt (2004), internal secular evolution of galaxy discs leads
to formation of the central mass concentrations of stars and gas in the
discs, generally called "pseudobulges"(PB). They look like classical,
merger-built bulges but that were made slowly out of disc gas. The
pseudobulges can be distinguished from classical bulges by their flatter
shapes, nearly exponential brightness profiles as well as starburst activity.
It should be acknowledged that past papers such as Simien \& de Vaucouleurs
(1986) confused pseudobulges with classical bulges and actually estimated
PB/D ratio.

  Kormendy et al. (2010) used HST archive images to measure the brightness
profiles of pseudobulges in M101 and NGC 6946. They showed that the above
galaxies contain only tiny pseudobulges that make up less than 3 percent
of the galaxy stellar mass. We kept this data as reference when visually assessing the proportion of a pseudobulge in the total luminosity of the galaxy. Taking into account the colour differences of a galaxy in the bands: $g, r, i, z, y$ in the Pan-STARRS1 survey, we estimated the relative luminosity of a  PB on scales: 5\%, 10\%, 15\%, and 20\%. Table 2 shows the result.  For 85\% face-on galaxies from our sample, the estimates of the relative luminosity of pseudobulge are 5--15\% with the mean $\langle PB/Tot\rangle= 11 $\%. This value almost coincides with the quantity $\langle (P)B/Tot\rangle=(13\pm3)$\% obtained by Davis et al. (2018) for Sc--Scd galaxies. 

There are more data on the decomposition of galaxies into a bulge and a disc in the literature. Some of them (Simard et al. 2011, Lackner \& Gunn 2012, and Lange et al. 2016) number hundreds of thousands of objects. However, we could not use these data, because decomposition algorithms split most of
our galaxies into separate nodes (then this is a sign that they have pseudobulges since star formation is often clumpy).

In general, it can be stated that the term ``bulgeless'' is fully justified in relation to spiral galaxies of the Sc--Sd types. 

\section{Nuclei as black hole candidates} Galaxies without any evidence of a bright bulge and face-on oriented are the most suitable objects for searching for unresolved nuclei in their central parts as candidates to black holes (BH). We noticed the presence of an unresolved nucleus in a galaxy using a simple scheme: ``2'' is a distinct star-shaped core, ``1'' is the central concentration of a not completely unresolved, but rather diffuse shape, ``0'' is the absence of any visible signs of a nucleus. We applied this classification both to the images of galaxies in five optical bands of Pan-STARRS1 and to the view of the central region of the galaxy in the 2MASS $K_s$- band. Statistics of the visibility of nuclei in 220 galaxies of our sample is represented by the matrix in Table 3.  The matrix of galaxy numbers has an asymmetric non-diagonal view with a shift towards a larger number of unresolved nuclei marked in the optical bands. This asymmetry is due to short galaxy exposures in the 2MASS survey. 

From this matrix, we have selected two extreme cases: ``2,2'' or ``0,0'' that mean presence or absence of unresolved nuclei signs according to the data from both surveys. The distributions of these galaxies by the absolute magnitude $M_B$ are presented in two panels of Fig. 6. As one can see, the galaxies with distinct signs of unresolved nuclei have much greater luminosity with the average absolute magnitude $\langle M_B\rangle_{22}=-20.73\pm0.08$, while the other category of galaxies is characterized by the average absolute magnitude $\langle M_B\rangle_{00}=-18.73\pm0.13$. This difference agrees well with the data obtained by Davis et al. (2018), according to which the BH mass and the stellar mass of the galaxy are related with the steep relation

$$\log M_{BH}\sim(3.0\pm0.5)\times \log M_*$$ with a scatter of 0.68 dex. Let us note that this relation was derived by Graham et al. (2018) for 74 spiral galaxies in the Virgo cluster, where the properties of the BH candidates can be dependent on the density of galaxy environment. Sanchez-Janssen et al. (2018) recently noticed such an effect. That is why the relation between $M_{BH}$ and $M_*$ can be less scattered for solitary late-type galaxies. 

 However, it should be noticed that classical bulges show a correlation between black hole mass and bulge mass but pseudobulges do not 
(Kormendy \& Ho, 2013). In fact, there is no correlation between pure-disk galaxies (e. g., disk mass) and black hole mass (see Figure 22 in the Kormendy \& Ho (2013)).

We did not make it our purpose to investigate the photometric properties of unresolved nuclei in our sample. However, comparing face-on bulgeless galaxies and the objects studied by Graham et al. (2018), we conclude that typical masses of BH candidates among our galaxies are in the range of $\sim(10^3-10^5) M_{\odot}$. 

According to Davis et al. (2017), BH masses show the closest correlation not with the stellar mass of a galaxy but with the spiral arm pitch angle $\phi$: 

$$\log M_{BH}\sim - (0.171\pm0.017)\times(|\phi^{\circ}|-15^{\circ}).$$ 

Our sample of bulgeless galaxies, whose spiral pattern is not distorted by the projection effect, provides an opportunity to test the degree of correlation of $M_{BH}$ with the pitch angle $\phi$ as well as with the presence of a bar in a galaxy. 

 Sometimes, central unresolved light sources are AGNs, not star clusters (e. g., NGC 4395, see Filippenko \& Ho (2003) and Kormendy \& Ho (2013)). 
Ho (2008) reviewed the subject of AGNs in galaxies such as those in our Table 1. Kormendy et al. (2010) provided black hole mass limits for several galaxies in Table 1. Interestingly, we found only a single galaxy, NGC~7798 = Mrk~332, among 220 face-on bulgeless galaxies that has a highly active nucleus. The nuclei of the rest bulgeless galaxies do not stand out with their strong emission. 
\begin{figure} \includegraphics[scale=0.7]{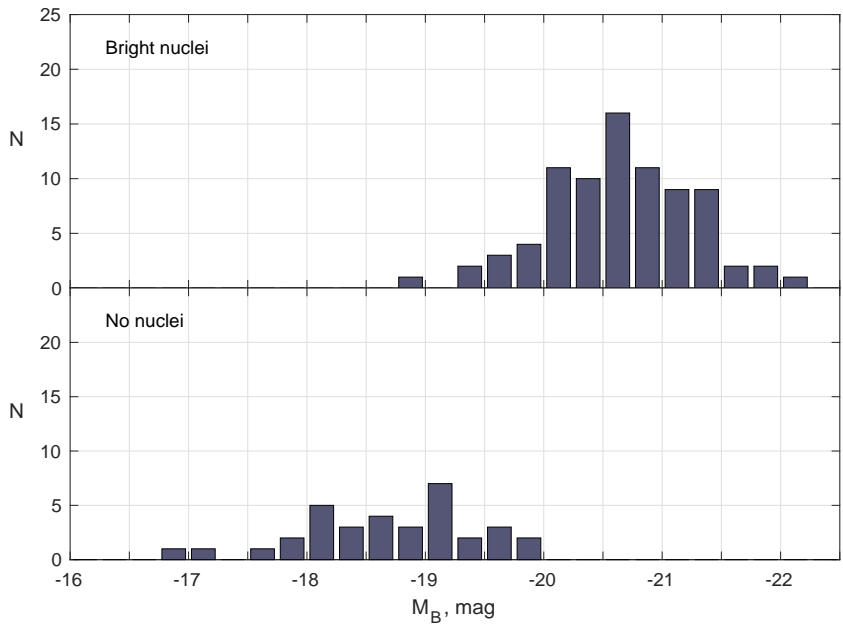} \caption{Distribution of face-on bulgeless galaxies with  signs of the presence of unresolved nucleus (the top panel) and without any (the bottom panel) by absolute magnitudes.} \end{figure}
The existing steep relation between the nucleus luminosity and the total stellar mass of a galaxy allows one to avoid rough errors in estimating the galaxy distance. For example, the Sc galaxy NGC~4136 with the radial velocity $V_{LG}=560$ km/s and the apparent magnitude $B=11.9^m$ has a distinct unresolved nucleus. With a Hubble distance of 8 Mpc, its absolute magnitude, $-17.7^m$, would not correspond to the presence of a noticeable nucleus (see Fig. 6). However, NGC~4136 is located in the ``Local Velocity Anomaly'' region known as Coma~I (Kashibadze et al. 2018). Being actually at   the Virgo cluster distance, 17 Mpc, this galaxy with $M_B=-19.3^m$ does not seem an exception to the general pattern. Thus, NGC~4136 is a possible new member of the peculiar association Coma~I numbering about 30 members. 

\section{Concluding remarks} Based on the data from Pan-STARRS1, SDSS, 2MASS, and HyperLEDA, we compiled a sample of face-on bulgeless galaxies. The sample includes 220 Sc--Scd--Sd galaxies distributed over the whole sky with the declinations DEC$>-30^{\circ}$. The galaxies have the angular diameters $d_{25}>0.8^{\prime}$, the typical inclination angle to the line of sight $i\sim23^{\circ}$, and the median radial velocity $\sim5000$ km/s. By its morphological composition and depth, the face-on bulgeless sample is close to the RFGC catalogue of flat galaxies; however, it contains a relatively smaller number of the Sd-type spirals than the sample of UF galaxies (Karachentseva et al. 2016). 

About a half of the face-on bulgeless galaxies (43\%) show the signs of bar-like structures, moreover, the bars can be found in the whole range of the absolute magnitudes $M_B$ from $-22^m$ to $-17^m$. Statistics of the spins of Sc--Scd--Sd face-on galaxies does not show any significant difference in the spin orientation of the spiral arms. A considerable amount of galaxies from our sample have slightly distorted peripheral spiral structures despite their apparent isolation. 

 According to our data, the pseudobulges of the Sc--Scd--Sd galaxies amount on average 11\% of the total luminosity (stellar mass) of galaxies. 
 Moreover, the largest representatives of this sample: M~101, IC~342, NGC~6946 are more robustly known to lack classical bulges according to Kormendy et al. ,2010. This suggests their classification as ``bulgeless discs''. 

Due to the negligible projection effect and slight internal extinction, our sample objects are quite suitable for studying the properties of central nuclei as possible BH candidates of moderate $(10^{3-5} M_{\odot}$) masses. Distinct unresolved nuclei in the $K_s$ band in 2MASS and in the $g, r, i, z, y$ bands in Pan-STARRS1 are observed in 40--60\% of galaxies from our list. The presence of {unresolved} nucleus closely correlates with the luminosity of a galaxy, indirectly favouring the known steep relation $M_{BH}\sim M^3_*$. 
The properties of BH candidates in face-on bulgeless galaxies are worth careful studying, especially with the database of Spitzer images. 

In our next paper we suppose to use the sample of face-on bulgeless galaxies together with the ultra-flat edge-on galaxy sample to elaborate the internal extinction effect in late-type galaxies as well as to determine star-formation rates in them. We also intend to estimate the occurrence of bulgeless galaxies in different cosmic structures in order to search for companions suitable to determine orbital masses from their motion. 

{\bf Acknowledgements} 

 The authors thank John Kormendy, the referee, for thorough examination of our manuscript and for very useful comments and suggestions to improve the text. We are also grateful to Dmitry I. Makarov and Olga G. Kashibadze for their special assistance in the work that we have undertaken and for  discussions.  The work was performed within the SAO RAS state assignment in the part "Conducting Fundamental Science Research".
The paper makes use of the data from the Pan-STARRS1, and the Two Micron All Sky Survey (2MASS) as well as from the HyperLEDA database (http://leda.univ-lyon1.fr). 

{\bf References} 

Abazajian K.N., Adelman-McCarthy J.K., Agueros M.A., et al., 2009, ApJS, 182, 54 (SDSS)

Chambers K.C., Magnier E.A., Metcalfe N., et al. 2016, arXiv:1612.05560 (Pan-STARRS1)

Davis B.L., Graham A.W., Cameron E., 2018, arXiv:1810.04888 

Davis B.L., Graham A.W., Seigar M.S., 2017, MNRAS, 471, 2187 

de Vaucouleurs, G., de Vaucouleurs, A., and Corwin, H. G. 1976, Second Reference Catalogue of Bright Galaxies, Austin: University of Texas Press (RC2) 

Filippenko, A. V., \& Ho, L. C. 2003, ApJ, 588, L1

Graham A.W., Soria R., Davis B.L., 2018, arXiv:1811.03232 

Heidmann J., Heidmann N., de Vaucouleurs G., 1972, MemRAS, 75,85 

Ho, L. C. 2008, ARA\&A, 46, 475

Jarrett, T.N., Chester, T., Cutri R. et al., 2000, AJ, 119, 2498 (2MASS)

Karachentsev I.D., Kaisina E.I., Kashibadze O.G., 2017, AJ, 153, 6 

Karachentsev I.D., Karachentseva V.E., Kudrya Y.N., 2016, AstBu, 71, 129  

Karachentsev I.D., Karachentseva V.E., Kudrya Y.N., et al., 1999, AstBu, 47, 5 (RFGC) 

Karachentsev I.D., Karachentseva V.E., Parnovsky S.L., 1993, Astron. Nachr. 314, 97 (FGC) 

Karachentseva V.E., Kudrya Y.N., Karachentsev I.D., Makarov D.I., Melnyk O.V., 2016, AstBu, 71, 1 (UF) 

Kashibadze O.G., Karachentsev I.D., Karachentseva V.E., 2018, AstBu, 73, 124 

Kim K., Oh S., Jeong H. et al., 2016, ApJS, 225, 6 

Kormendy, J. 2013, in Canary Islands Winter School of Astrophysics, Volume XXIII, Secular
  Evolution of Galaxies, ed. J. Falc\'{o}n-Barroso \& J. H. Knapen (Cambridge: Cambridge
  University Press), 1

Kormendy, J., \& Ho, L. C. 2013, ARA\&A, 51, 511

Kormendy J., Drory N., Bender R., Cornell M., 2010, ApJ, 723, 54 

Kormendy, J. \& Kennicutt, R.C., 2004, ARA\&A, 42, 603

Kormendy, J. 1993, in IAU Symposium 153, Galactic Bulges, ed. H. Dejonghe \& H. J. Habing
  (Dordrecht: Kluwer), 209

Kormendy, J. 1982, in Twelfth Advanced Course of the Swiss Society of Astronomy and
 Astrophysics, Morphology and Dynamics of Galaxies, ed. L. Martinet \& M. Mayor (Sauverny: 
 Geneva Observatory), 113

Kormendy J., 1979, ApJ, 227, 714

Lackner C.N., Gunn J.E., 2012, MNRAS, 421,2277 

Lange R., Moffett A.J., Driver S.P., et al, 2016, MNRAS, 462, 1470 

Makarov D., Prugniel P., Terekhova N., et al. 2014, A \& A, 570A, 13 (HyperLEDA) 

Melnyk O.V., Karachentseva V.E., Karachentsev I.D., 2017, AstBu, 72, 1  

Oohama N., Okamura S., Fukugita M., et al. 2009, ApJ, 705, 245 

Parnovsky S.L., Karachentsev I.D., Karachentseva V.E., 1994, MNRAS, 268, 665 

Sanchez-Janssen R., Cote P., Ferrarese L., et al., 2018, arXiv:1812.01019 

Schlegel D.J., Finkbeiner D.P., \& Davis M., 1998, ApJ, 500, 525 

Simard L., Mendel J.T., Patton D.R., et al. 2011, ApJS, 196, 11 

Simien F., de Vaucouleurs G., 1986, ApJ, 302, 517

\end{document}